\documentclass[11pt]{article}
\usepackage[margin=1in]{geometry}
\usepackage{amsmath,amssymb,amsthm}
\usepackage{booktabs}
\usepackage{array}
\newcolumntype{L}[1]{>{\raggedright\arraybackslash}p{#1}}
\usepackage[authoryear]{natbib}
\usepackage[colorlinks=true, citecolor=blue, linkcolor=blue, urlcolor=blue]{hyperref}

\newtheorem{proposition}{Proposition}
\theoremstyle{remark}
\newtheorem*{remark}{Remark}

\newcommand{\supp}{\operatorname{supp}}
\newcommand{\rank}{\operatorname{rank}}
\title{Decorrelation Before Variable Selection:\\ Left and Right Transforms and Their Trade-Off}
\author{Mahdi Nouraie$^{a,}$\thanks{Corresponding author. Email: \texttt{soheil.nouraie@mq.edu.au}} \and Samuel Muller$^{a,b}$\\[1ex]
\small $^{a}$School of Mathematical and Physical Sciences, Macquarie University, Sydney, NSW 2109, Australia\\
\small $^{b}$School of Mathematics and Statistics, The University of Sydney, Sydney, NSW 2006, Australia}
\date{}

\begin{document}
\maketitle

\begin{abstract}
Decorrelation has two forms. A left transform keeps coefficients and changes noise; a right transform keeps noise
and changes coefficients. Cholesky whitening alone keeps noise variables inactive when true ones come first;
decorrelation preserving coefficients and noise loses information.
\end{abstract}


\section{Introduction}

Consider the linear model without intercept
\begin{equation}\label{eq:model}
  y = X\beta + \varepsilon, \qquad \operatorname{Cov}(\varepsilon) = \sigma^2 I_n,
\end{equation}
with $X \in \mathbb{R}^{n\times p}$ column-centred, so that $\rank(X) \le n-1$, and $\beta$ sparse. We call
$S = \{j : \beta_j \ne 0\}$ the set of \emph{true variables} and the others \emph{noise variables}, and we say
that a variable is \emph{active} in a model if its coefficient in that model is non-zero, and \emph{inactive}
otherwise.

Correlation among the columns of $X$ can violate the irrepresentable condition
\citep{zhao2006} for the Lasso \citep{tibshirani1996}, weaken marginal screening \citep{fan2008} and make selection unstable
\citep[see the discussion in][]{meinshausen2010}. A common solution in variable selection and screening is to transform the data
so that the columns become uncorrelated, or nearly so, before selecting. Linear transforms come in two
kinds, left (row) and right (column) transforms, a distinction drawn by \citet{kelner2021} for
preconditioning in the space of samples and of parameters:
\[
  Fy = (FX)\beta + F\varepsilon, \quad \operatorname{Cov}(F\varepsilon) = \sigma^2 FF^\top;
  \qquad\qquad
  y = (XG)\,\gamma + \varepsilon, \quad \gamma = G^{-1}\beta ,
\]
with $F \in \mathbb{R}^{m\times n}$ and an invertible $G \in \mathbb{R}^{p\times p}$. A left transform keeps the
coefficients $\beta$ but transforms the noise (for whitening, the two views coincide up to which parametrisation
is penalised; end of Section~3). A right transform keeps the noise and, for fixed $G$, each row
depends only on its own observation, but the coefficients, and hence the meaning of the columns, change.
\citet{wauthier2013} compare left preconditioners for the Lasso in a common framework that allows different
matrices for $X$ and $y$; we apply the same $F$ to both, so that model~\eqref{eq:model} is preserved.

We show that existing methods are, at their core, left transforms (preconditioning, DECO, HOLP screening,
Cholesky-based penalisation), right transforms (whitening, Gram--Schmidt orthogonalisation,
weighted Lassos) or a combination of both, and we give two results that explain their trade-offs.
Proposition~\ref{prop:chol} shows that, among right whitenings, Cholesky alone keeps noise variables
inactive whenever the true variables come first, whatever their number and coefficients.
Propositions~\ref{prop:imposs} and~\ref{prop:proj} show that no fixed linear transform decorrelates the
columns while keeping both the coefficients and white noise, unless it discards information; the whitened estimator that underlies fixed-X knockoffs is such a reduced
model.

\section{Left transforms}

Let $X = UDV^\top$ be the compact singular value decomposition, with $D$ the diagonal matrix of the
$\rank(X)$ non-zero singular values. Since $X$ is centred, $XX^\top$ is singular. \citet{jia2015}, \citet{wang2016deco}
and \citet{wang2016holp} write their transforms or estimators with $(XX^\top)^{-1}$ or its square root, which requires $XX^\top$ to be
invertible. \citet{wang2016deco} note that this fails after standardisation and use a ridge version, and they also use the
pseudo-inverse when $p \le n$; \citet{wang2016holp} also propose a ridge version, recommended when $XX^\top$ is close to
degeneracy. For centred $X$ we read $(XX^\top)^{-1}$ as the pseudo-inverse $(XX^\top)^{+} = UD^{-2}U^\top$.
Ridge parameters $r$ below are positive.

\paragraph{Puffer transform \citep{jia2015}.} Take $F = UD^{-1}U^\top$. Then
\[
  FX = UV^\top ,\qquad (FX)^\top FX = VV^\top, \qquad \operatorname{Cov}(F\varepsilon) = \sigma^2\, U D^{-2} U^\top .
\]
If $\rank(X) = p$, then $VV^\top = I_p$ and the columns of $FX$ are exactly orthonormal; otherwise,
in particular if $p \ge n$, $VV^\top$ is only a projection and the columns are in general not orthogonal.
\citet{jia2015} show that $FX$ can satisfy the irrepresentable condition when $X$ does not, so the
preconditioned Lasso is sign consistent in a wider range of settings. The price is the noise: $F$ inflates the smallest
singular values of $X$ to one, which also inflates and correlates the errors.

\paragraph{DECO \citep{wang2016deco}.} Aimed mainly at $p \gg n$, DECO uses $F = (XX^\top/p)^{-1/2} = \sqrt{p}\,UD^{-1}U^\top$,
the Puffer transform up to the factor $\sqrt{p}$; because $XX^\top$ has rank at most $n-1$ after
standardisation, its algorithm uses $\sqrt{p}\,(XX^\top + rI_n)^{-1/2}$. After decorrelation, the variables are
split across machines and a sparse regression is fitted to each block. For both transforms, each transformed
row typically mixes all $n$ observations, so a subsample of transformed rows has correlated noise.

\paragraph{HOLP \citep{wang2016holp,wang2021holp}.} HOLP ranks the variables by the absolute values of the entries of
$\hat\beta^{\mathrm{HOLP}} = X^\top (XX^\top)^{-1} y$. With the Puffer $F$,
$(FX)^\top (Fy) = X^\top U D^{-2} U^\top y = X^\top (XX^\top)^{+} y$,
so HOLP is marginal screening, as in SIS \citep{fan2008}, applied after left decorrelation.
Ridge-HOLP uses $X^\top (XX^\top + rI_n)^{-1} y = (F_rX)^\top F_r y$ with $F_r = (XX^\top + rI_n)^{-1/2}$, which is also
the ridge regression estimate \citep{hoerl1970}; Air-HOLP \citep{joudah2025} chooses $r$ adaptively.

\paragraph{Cholesky as a left transform.} If $\rank(X) = p$, write the thin QR factorisation $X = QR$, with $R$
upper triangular with positive diagonal, so that $X^\top X = R^\top R$ and $R$ is the Cholesky factor. Taking
$F = Q^\top$ gives
\[
  Q^\top y = R\beta + Q^\top\varepsilon, \qquad \operatorname{Cov}(Q^\top\varepsilon) = \sigma^2 I_p ,
\]
and $\|y - X\beta\|^2 = \|Q^\top y - R\beta\|^2 + \|(I_n - QQ^\top)y\|^2$, so the least-squares estimate and the Lasso
path are unchanged (with the loss scaled by the same constant). The Cholesky matrix penalisation of \citet{nghiem2022} has this form:
for sliced inverse regression, it solves $\Sigma\beta = \eta$ through $\Sigma = LL^\top$ by a Lasso on
$L^\top\beta \approx L^{-1}\eta$; with $\Sigma = X^\top X$ and $\eta = X^\top y$, this is the Lasso on $(Q^\top y, R)$.
The transform keeps $\beta$ and white noise but does not decorrelate, since $R^\top R = X^\top X$.

\section{Right transforms}

A right transform $G$ \emph{whitens} $X$ if $G^\top X^\top X G = I_p$; this requires $\rank(X) = p$, hence $p \le n-1$ because $X$ is centred.
Writing $M = G^{-1}$, the condition is $M^\top M = X^\top X$, and the transformed coefficient is $\gamma = M\beta$.
\citet{kessy2018} compare the standard choices:
\begin{itemize}
  \item \textbf{ZCA} (zero-phase component analysis, or Mahalanobis whitening): $G = (X^\top X)^{-1/2}$. The
  whitened data stay as close as possible to the original data. \citet{zhu2021} apply ZCA-type whitening,
  with an estimated correlation matrix of the predictors, before a Lasso-type selection, noting that, unlike
  left preconditioning, the error term is unchanged.
  \item \textbf{PCA} (principal component analysis): $G = VD^{-1}$, so $XG = U$. Each new column is in general
  a mix of all original variables.
  \item \textbf{Cholesky:} $G = R^{-1}$, so $XG = Q$. Computationally this is Gram--Schmidt
  orthogonalisation: column $j$ of $Q$ is the part of $X_j$ not explained by $X_1,\dots,X_{j-1}$, normalised to
  unit length, so the result depends on the ordering of the columns.
\end{itemize}

A desirable property for selection is that a right transform does not move signal into columns that come
later in the ordering than the true variables. The next result shows that this property singles out
Cholesky.

\begin{proposition}\label{prop:chol}
Assume $X^\top X$ is positive definite, let $M \in \mathbb{R}^{p\times p}$ satisfy $M^\top M = X^\top X$, and fix the
ordering of the columns. The following are equivalent:
\begin{enumerate}
  \item[(a)] for every $k$ and every $\beta$ with $\supp(\beta) \subseteq \{1,\dots,k\}$, we have
  $\supp(M\beta) \subseteq \{1,\dots,k\}$, where $\supp(v) = \{j : v_j \ne 0\}$;
  \item[(b)] $M$ is upper triangular;
  \item[(c)] $M = \Delta R$, where $\Delta$ is diagonal with entries $\pm 1$ and $R$ is the Cholesky factor of $X^\top X$.
\end{enumerate}
\end{proposition}

\begin{proof}
(a)$\Rightarrow$(b): take $\beta = e_j$, whose support lies in $\{1,\dots,j\}$. Then $Me_j$, the $j$th
column of $M$, is supported on $\{1,\dots,j\}$, so $M_{ij} = 0$ for $i > j$.
(b)$\Rightarrow$(a): if $M$ is upper triangular, then $(M\beta)_i = \sum_{l \ge i} M_{il}\beta_l$, which vanishes for
$i > k$ whenever $\beta_l = 0$ for $l > k$.
(b)$\Rightarrow$(c): $M$ is invertible because $X^\top X$ is, so its diagonal entries are non-zero. Let
$\Delta = \operatorname{diag}(\operatorname{sign}(M_{11}),\dots,\operatorname{sign}(M_{pp}))$. Then $\Delta M$ is upper triangular with positive
diagonal and $(\Delta M)^\top(\Delta M) = X^\top X$, so $\Delta M = R$ by uniqueness of the Cholesky factor, and $M = \Delta R$.
(c)$\Rightarrow$(b) is immediate.
\end{proof}

The equivalence of (b) and (c) is the uniqueness of the Cholesky factor, and \citet[Section~6.5]{kessy2018}
characterise Cholesky whitening by the analogous triangularity of the cross-product $(XG)^\top X = M$.
Their whitening matrix acts on a column vector, so it corresponds to $G^\top$ here; they use the
Cholesky factor of the precision matrix, which makes $M$ lower triangular and amounts to the reversed
ordering. \citet{kelner2021} precondition the Lasso on the right with a sparse block Cholesky factorisation, so
that, for low-treewidth designs, sparse coefficient vectors stay sparse; condition (a) asks for more, that the support stay within the
leading coordinates. Triangularity itself is classical, since Gram--Schmidt orthogonalisation is sequential regression
\citep[Section~3.2.3]{hastie2009}; what we add, to our knowledge, is condition (a), which reads it as a
selection property and shows that Cholesky is the only whitening with it.
The quantifier ``for every $k$'' matters: for a single known $k$, any whitening with $M$ block upper triangular
at $k$ also satisfies (a) for that $k$.

Here, active and inactive refer to the coefficients $\gamma = M\beta$ of the transformed model, not to an
estimate; a noise variable with $\gamma_j \ne 0$ is a correct selection in the transformed model but a false one in
terms of the original variables. Thus, if the true variables come first, Cholesky is, up to signs, the only
whitening that keeps the noise variables inactive, $\gamma_j = 0$ for $j > |S|$, for every number of true
variables and every coefficient vector on them. Two limitations remain.
\begin{itemize}
  \item True variables may become weak. Since $\gamma_j = \sum_{l \ge j} R_{jl}\beta_l$, a true variable that is well
  explained by earlier ones has a small $R_{jj}$, and $\gamma_j$ can even vanish through cancellation. No
  information is lost, since the transform is invertible; the shared signal is carried by earlier columns.
  \item The ordering must be right. A noise variable placed before a true variable with which it is
  correlated typically absorbs some of its signal. The requirement that each true variable precede the noise
  variables correlated with it, which \citet[Section~2.3]{nouraie2026} use for consistent selection, does not keep them inactive. Take true variables $T_1, T_2$ and a
  noise variable $N$ in the ordering $(T_1, N, T_2)$, with correlations between the columns $\rho(T_1,N) = 0.6$,
  $\rho(T_1,T_2) = 0.5$, $\rho(N,T_2) = 0$ and $\beta = (1,0,1)$. For unit-norm columns, the Cholesky factor of the correlation matrix gives
  $\gamma = (1.5,\,-0.375,\,0.781)$, so $N$ picks up signal from $T_2$ through $T_1$ and is active. This
  requirement does ensure $\gamma_N = 0$ when groups of correlated variables are mutually uncorrelated and,
  within each group, every noise variable is correlated with every true variable.
\end{itemize}

When $XG$ has orthonormal columns, the Lasso on $XG$ is soft-thresholding: minimising
$\tfrac12\|y - XG\gamma\|^2 + \lambda\|\gamma\|_1$ gives $\hat\gamma_j = \operatorname{sign}(z_j)(|z_j| - \lambda)_+$, with
$z = (XG)^\top y = M\hat\beta^{\mathrm{OLS}}$ and $\operatorname{Cov}(z) = \sigma^2 I_p$. So every right whitening thresholds a different
linear image of the least-squares estimate, ZCA, PCA and Cholesky differing only in $M$, whereas Puffer,
for $n > p$ and full-rank $X$, thresholds $\hat\beta^{\mathrm{OLS}}$ itself \citep{rohe2015}, with correlated noise \citep{jia2015}. Equivalently, a whitened right-transform Lasso is a
left transform $F = (XG)^\top$, which gives $Fy = M\beta + F\varepsilon$ with $FX = M$ and $\operatorname{Cov}(F\varepsilon) = \sigma^2 I_p$,
followed by a Lasso that penalises $M\beta$ instead of $\beta$. For Cholesky, $F = Q^\top$ and $M = R$, so the left and right versions use
the same statistic $Q^\top y$ and differ only in which parametrisation is penalised.

\section{Combining left and right transforms}\label{sec:dvs}

A method can use one kind of transform for ranking and the other for selection. DVS \citep{nouraie2026}
ranks the variables by Air-HOLP, applies Gram--Schmidt to the columns of $X$ in this order, and runs stability
selection \citep{meinshausen2010} with the Lasso on $Q$; since Ridge-HOLP is screening after a left transform,
it uses a left transform for \emph{ranking} and a right transform for \emph{selection}. Two general points
apply to any such combination. First, the ordering uses $y$, so the guarantees above hold for a given ordering.
Second, the decorrelation is exact only on the full sample: a subsample $Q_I$ of $\lfloor n/2 \rfloor$ rows (index set $I$), as in
stability selection, has $\mathbb{E}[Q_I^\top Q_I] = (|I|/n)\, I_d$ under uniform subsampling, with $d \le \rank(X)$
orthogonalised columns, but each realisation deviates from this, and $Q_I^\top Q_I$ is singular once $d > \lfloor n/2 \rfloor$.

\begin{remark}
When $p > \rank(X)$, at most $\rank(X)$ columns can be orthogonalised. Once the columns already processed
span the column space of $X$, every further Gram--Schmidt residual is zero in exact arithmetic; numerically,
the normalised residuals are rounding error, giving meaningless unit-norm columns that the Lasso can still
select. Orthogonalising only the top $d \le \rank(X)$ ranked columns, as screening would, avoids this and, for
$d \le \lfloor n/2 \rfloor$, typically keeps $Q_I^\top Q_I$ non-singular, provided these columns are linearly independent; the price is that the true variables must
rank in the top $d$.
\end{remark}

\section{The trade-off}\label{sec:trade}

Assume here that $X^\top X$ is invertible, so that exact decorrelation is possible; this requires $p \le n-1$ ($X$ is centred),
so the results below do not cover the $p \ge n$ regime that motivates screening. A decorrelating transform
would ideally achieve all three of the following:
\begin{enumerate}
  \item[(i)] uncorrelated columns: the transformed Gram matrix is diagonal;
  \item[(ii)] unchanged meaning: the coefficients are $\beta$, up to a fixed diagonal rescaling and relabelling;
  \item[(iii)] unchanged noise: the transformed noise has covariance $\sigma^2 I$.
\end{enumerate}

\begin{proposition}\label{prop:imposs}
Let $X^\top X$ be invertible and not diagonal, and consider a fixed linear transform, that is, one that does not depend on $y$,
$(y, X) \mapsto (Fy, FXG)$ with $F \in \mathbb{R}^{m\times n}$ and an invertible $G \in \mathbb{R}^{p\times p}$. If it satisfies (i), (ii) for every $\beta$,
and (iii), then $m < n$, that is, the data are reduced to fewer than $n$ linear combinations of the observations.
\end{proposition}

\begin{proof}
Writing the reduced model as $Fy = (FXG)(G^{-1}\beta) + F\varepsilon$ and applying (ii) to $\beta = e_j$ gives $G^{-1}e_j = c_j e_{\pi(j)}$ with $c_j \ne 0$, for $j = 1,\dots,p$; since $G^{-1}$ is
invertible, $\pi$ is a permutation, so $G$ is a scaled permutation. Condition (iii) means $FF^\top = I_m$, hence $m \le n$.
If $m = n$, then $F$ is orthogonal, so the transformed Gram matrix is $G^\top X^\top X G$, which, $G$ being a scaled
permutation, is diagonal only if $X^\top X$ is. Hence (i) fails unless $m < n$.
\end{proof}

To our knowledge, Proposition~\ref{prop:imposs} and Proposition~\ref{prop:proj} below have not been stated in this form.
Any transform in this class that achieves (i) must give up (ii), give up (iii), or reduce the data:
Cholesky as a right transform gives (i) and (iii); Puffer gives (i) and (ii), and when $\rank(X) = p$, $FX = UV^\top$
is exactly the ZCA-whitened design, so the two differ only in whether $\beta$ or the noise is kept; Cholesky as a
left transform gives (ii) and (iii) with $m = p < n$, but not (i). Achieving all three requires $G$ to be a scaled
permutation and $F = W^\top$ with $W \in \mathbb{R}^{n\times m}$ having orthonormal columns and $m < n$.

\begin{proposition}\label{prop:proj}
Let $W \in \mathbb{R}^{n\times m}$ have orthonormal columns and suppose the columns of $W^\top X$ are orthogonal,
with $X^\top W W^\top X = \Lambda = \operatorname{diag}(\mu_1,\dots,\mu_p)$ and all $\mu_j > 0$. Then the reduced model
$W^\top y = (W^\top X)\beta + W^\top\varepsilon$ satisfies (i), (ii) and (iii), and
\[
  \Lambda \preceq X^\top X, \qquad\text{hence}\qquad \frac{\sigma^2}{\mu_j} \;\ge\; \sigma^2\big[(X^\top X)^{-1}\big]_{jj}.
\]
\end{proposition}

\begin{proof}
$\operatorname{Cov}(W^\top\varepsilon) = \sigma^2 W^\top W = \sigma^2 I_m$, the coefficients are $\beta$, and the Gram matrix is $\Lambda$.
Since $I_n - WW^\top \succeq 0$, we get $X^\top X - \Lambda = X^\top (I_n - WW^\top) X \succeq 0$. Both $\Lambda$ and $X^\top X$ are positive
definite, so inverting reverses the order, $\Lambda^{-1} \succeq (X^\top X)^{-1}$, and comparing diagonal entries gives the bound.
\end{proof}

The left-hand side of the bound is the variance of the least-squares estimate of $\beta_j$ in the reduced
model; the right-hand side is that from the full data, the Gauss--Markov bound for linear unbiased
estimators. If $X^\top X$ is not diagonal then $\Lambda \ne X^\top X$, so $\Lambda^{-1} - (X^\top X)^{-1}$ is non-zero and positive
semidefinite, hence has positive trace, and for at least one $j$ the variance $\sigma^2/\mu_j$ strictly exceeds the
full-data variance: the reduction loses information. Under strong correlation $[(X^\top X)^{-1}]_{jj}$ is large for the variables involved, so
$\mu_j \le 1/[(X^\top X)^{-1}]_{jj}$ is small for them. By contrast, the left Cholesky transform reduces the data to $p$ dimensions
without changing the least-squares estimate or its variance, because it does not decorrelate: the loss comes from (i), not from the reduction itself.

Such a $W$ need not exist. With $m = p$, one construction is $W = X(X^\top X)^{-1}\Lambda^{1/2} + Z$, where the
columns of $Z$ are orthogonal to those of $X$ and $Z^\top Z = I_p - \Lambda^{1/2}(X^\top X)^{-1}\Lambda^{1/2}$. This needs
$\Lambda \preceq X^\top X$ and $\rank(X^\top X - \Lambda) \le n - p$; for example, $\Lambda = \lambda_{\min}(X^\top X)\, I_p$ works whenever
$n \ge 2p-1$. For this $W$, $\Lambda^{-1/2}W^\top y = \hat\beta^{\mathrm{OLS}} + \Lambda^{-1/2}Z^\top\varepsilon$, the full-data least-squares
estimate plus extra noise. Under Gaussian errors, this has the same distribution as the whitened estimator of
\citet{li2021}, with covariance $\sigma^2\Lambda^{-1}$ (they add user-generated noise where we use the residual space):
their requirement that a diagonal $\Omega$ satisfy $\Omega \succeq (X^\top X)^{-1}$ is our $\Lambda \preceq X^\top X$ with $\Omega = \Lambda^{-1}$, and
they show that the price in variance can be very high when the eigenstructure of $X^\top X$ is unfavourable.

The fixed-X knockoff construction \citep{barber2015} supplies such a $W$; \citet{li2021} show that the fixed-X
knockoff filter is equivalent to inference on such a whitened estimator, built from the contrasts
$X_j - \tilde X_j$ \citep[eq.~(5)]{li2021}; the filter can also use $(X + \tilde X)^\top y$, which, under Gaussian errors, is independent of this
estimator, to rank the variables, together with the magnitude of the whitened estimator, so only the estimator it tests with is reduced. With columns normalised to unit
length, $\Sigma = X^\top X$ and $n \ge 2p$, the knockoffs exist when $0 \preceq \operatorname{diag}(s) \preceq 2\Sigma$; when all $s_j > 0$,
the contrasts $X_j - \tilde X_j$ are mutually orthogonal with squared norms $2s_j$, and taking $W$ with columns
$(X_j - \tilde X_j)/\sqrt{2s_j}$ gives $\Lambda = \operatorname{diag}(s)/2$. The constraint $\Lambda \preceq X^\top X$ is then exactly
$\operatorname{diag}(s) \preceq 2\Sigma$, so the $s_j$ must be small under strong correlation, one way to see why knockoffs
lose power for highly correlated variables. \citet{barber2015} recommend choosing the $s_j$ as large as possible,
whereas \citet{spector2022} show for model-X knockoffs that pushing $s$ to the boundary can itself reduce power, and propose
alternatives within the same constraint that are also more powerful in fixed-X simulations.

\paragraph{A meaning-preserving alternative.} Proposition~\ref{prop:imposs} also shows what remains when (ii)
and (iii) are both required without reducing the data: $G$ must be a scaled permutation, so the columns can
only be rescaled. A weighted Lasso, such as the adaptive Lasso \citep{zou2006}, is exactly of this form. With
positive penalty factors $w_j$, minimising $\|y - X\beta\|^2 + \lambda\sum_j w_j|\beta_j|$ is the ordinary Lasso applied to
$XG$ with $G = \operatorname{diag}(1/w_1,\dots,1/w_p)$, whose coefficients $\gamma_j = w_j\beta_j$ keep their meaning; a zero weight
leaves that variable unpenalised (the objective then has no such $G$). The Stable Lasso \citep{nouraie2025stable} sets $w_j$ to an increasing function
of the Air-HOLP rank of variable $j$ (rank 1 being the most relevant), so that, as in DVS, a left transform supplies the ranking. When the weights
depend on $y$, the argument applies for given weights. Such methods give up (i), but by the proof of
Proposition~\ref{prop:imposs}, rescaling, up to a square orthogonal $F$ and a reordering of the columns, which leave the Lasso
unchanged, is the only option that keeps both the meaning of the coefficients and white noise without reducing
the data. Like Puffer, their gains come from changing how the Lasso behaves, not from new information.

\section{Discussion}\label{sec:disc}

No transformation of the data, linear or not, can create information about $\beta$; under Gaussian errors and
with $X^\top X$ invertible, any unbiased estimate of $\beta_j$ has variance at least $\sigma^2[(X^\top X)^{-1}]_{jj}$, however the
data are transformed. By Propositions~\ref{prop:imposs} and~\ref{prop:proj}, achieving (i)--(iii) together loses
information unless the columns of $X$ are already orthogonal. This is a statement about information, not about
procedures: because the Lasso is nonlinear and ignores the noise covariance, a transform can still change how
well it selects, as the Puffer results of \citet{jia2015} show; conversely, \citet{kelner2021} construct Gaussian designs of
high treewidth for which no right-preconditioned $\ell_1$ program recovers sparse vectors from fewer than polynomially many samples in the treewidth. Table~\ref{tab:pay} summarises which property each method gives up.

\begin{table}[t]
\centering
\caption{Properties of decorrelating transforms.}\label{tab:pay}
\small
\begin{tabular}{@{}L{2.6cm}L{0.9cm}L{3.0cm}L{2.6cm}L{2.2cm}L{1.8cm}@{}}
\toprule
Method & Side & Uncorrelated columns & Keeps $\beta$ & Noise stays white & Needs ordering \\
\midrule
Puffer, DECO & left & exactly if $\rank(X) = p$ (Puffer); otherwise not in general & yes & no & no \\
Cholesky ($Q^\top$)$^{\dagger}$ & left & no & yes & yes & no \\
Projection ($W^\top$)$^{*}$ & left & exactly & yes & yes & no \\
ZCA, PCA & right & exactly; needs $\rank(X) = p$ & no & yes & no \\
Cholesky (Gram--Schmidt) & right & exactly on the full sample, for at most $\rank(X)$ columns & no; noise variables stay inactive if true ones come first & yes, for a fixed ordering & yes \\
Weighted Lasso & right (diagonal) & no & yes, up to scaling & yes, for fixed weights & no \\
\bottomrule
\end{tabular}

\smallskip
\footnotesize $^{\dagger}$Reduces the data to $p$ dimensions without changing the least-squares estimate or the Lasso path; needs $\rank(X) = p$; used by \citet{nghiem2022}.
$^{*}$Reduces the data and loses information; exists, for example, when $\rank(X) = p$ and $n \ge 2p-1$; never, with all
$\mu_j > 0$, when $\rank(X) < p$.
\end{table}

In practice, the choice depends on what the analysis needs. If the goal is to select among the original
variables and interpret their coefficients, a method that keeps $\beta$, such as a weighted Lasso, is the natural
choice, and the correlation has to be handled by the selection method itself. If decorrelated columns are
worth losing that interpretation, Cholesky is the right transform to use, but its results are only as good as the
ordering, and its decorrelation is exact only on the full sample. Methods that keep both $\beta$ and white noise
while decorrelating exist only by losing information, as Proposition~\ref{prop:proj} shows; the
whitened estimator behind fixed-X knockoffs is an example.

\section*{Funding}
Mahdi Nouraie began this work during his PhD, supported by the Macquarie University Research Excellence
Scholarship (20213605), and continued it with support from the Australian Research Council Discovery Project
Grant (DP260100348), which also supported Samuel Muller.

\section*{Data availability}
No data were used for the research described in this article.

\section*{Declaration of generative AI and AI-assisted technologies in the manuscript preparation process}
During the preparation of this work the authors used Claude (Anthropic) to help edit the text and to check
the derivations. After using this tool, the authors reviewed and edited the content as needed
and take full responsibility for the content of the published article.

\bibliographystyle{plainnat}
\bibliography{refs}

\end{document}